\documentclass[prb,floatfix,showpacs,amsfonts,amssymb,twocolumn]{revtex4}
\usepackage{graphics}
\usepackage{dcolumn}
\usepackage{bm}
\usepackage{subfigure}
\usepackage{epsfig}

\usepackage{amsmath}
\usepackage{amsfonts}
\usepackage{amssymb}

\usepackage[colorlinks=true,linkcolor=blue]{hyperref}%
\expandafter\ifx\csname package@font\endcsname\relax\else
 \expandafter\expandafter
 \expandafter\usepackage
 \expandafter\expandafter
 \expandafter{\csname package@font\endcsname}%
\fi
\begin{document}

\title{Skyrmion confinement in ultrathin film nanostructures in the presence of Dzyaloshinskii-Moriya interaction}

\author{S.~Rohart}
\email{rohart@lps.u-psud.fr}
\affiliation{Laboratoire de Physique des Solides, Universit{\'{e}} Paris-Sud,
CNRS UMR 8502, F-91405 Orsay Cedex, France}
\author{A.~Thiaville}
\affiliation{Laboratoire de Physique des Solides, Universit{\'{e}} Paris-Sud,
CNRS UMR 8502, F-91405 Orsay Cedex, France}

\date{\today}

\begin{abstract}
We study the modification of micromagnetic configurations in nanostructures, due to the Dzyaloshinskii-Moriya interaction (DMI) that appear at the interface of an ultrathin film.
We show that this interaction leads to new micromagnetic boundary conditions that bend the magnetization at the edges.
We explore several cases of ultrathin film nanostructures that allow analytical calculations
(1D systems, domain walls, cycloids and skyrmions), compare with fully numerical calculations,
and show that a good physical understanding of this new type of micromagnetics can be reached. We particularly focus on skyrmions confined in circular nanodots and show that edges allow for the isolation of single skyrmions for a large range of the DMI parameter.
\end{abstract}



\maketitle

\section{Introduction}

Recent observations of chiral structures in magnetic thin films~\cite{bode2007,heinze2011,yu2010,yu2012,chen2013}
have raised a great interest for the Dzyaloshinskii-Moriya interaction
(DMI)~\cite{dzialo1957,dzyalo1964,dzyalo1965,moriya1960},
as it favors magnetization rotations with a fixed chirality\cite{dzyalo1965,bogdanov1989,bogdanov2001}.
This coupling originates from the combination of low structural symmetry and large spin-orbit coupling.
It has been first proposed in bulk materials lacking space inversion symmetry~\cite{moriya1960} but it also
exists at the interface between a magnetic film and a high spin-orbit coupling adjacent layer.~\cite{fert1980,fert1990}
The most striking phenomenon induced by DMI is the formation of skyrmion networks~\cite{bogdanov1994,rossler2006,yu2010,heinze2011}, but
its influence on domain walls~\cite{heide2008,thiaville2012,chen2013,emori2013,ryu2013} is also at the origin of interesting properties such
as increased domain wall velocity versus magnetic field. Recently, interest has also been devoted to isolated skyrmions, which can be nucleated as a metastable state in thin films\cite{romming2013,sampaio2013}, opening a path to new concepts of magnetic memories based on skyrmion motion in nanotracks\cite{fert2013,sampaio2013}.

While extensive work has already been performed on the influence of DMI on micromagnetism for infinite samples~\cite{dzyalo1964,dzyalo1965,bogdanov1994,bogdanov2001,rossler2006,heide2008,ezawa2010,kiselev2011prl,kiselev2011,thiaville2012}, no description is available for nanostructures, which is the aim of the present work.
We show that in nanostructures, DMI leads to a new form of micromagnetic boundary conditions that should be implemented in micromagnetic numerical solvers.
We describe several cases with analytical solutions that provide tests for numerical codes, and help
to get a physical feeling of the effects of this interaction. We particularly focus on the problem of skyrmions trapped in nanodots. Using simple physical arguments based on the micromagnetic length scales, we discuss the different states that are obtained. This should help future studies to design new memories based on skyrmion motion\cite{fert2013}. As most of the recent advances in this field toward application in spintronics devices have been obtained for ultrathin films\cite{thiaville2012,emori2013,ryu2013,romming2013,sampaio2013,fert2013}, we restrict our study to this case, using the interfacial DMI coupling described by A. Fert\cite{fert1990} and using a 2D formulation, where any variation along the film normal is neglected.


\section{Micromagnetic framework}


The Dzyaloshinskii-Moriya interaction has been introduced \textbf{in} an atomic description
as~\cite{dzialo1957,moriya1960,dzyalo1965}
\begin{equation}\label{eq_DMI_atomique}
    E_{\mathrm{DM}} = \sum_{\langle i,j\rangle}\vec{d}_{ij}.\left(\vec{S}_i\times\vec{S}_j\right)
\end{equation}
where $\vec{d}_{ij}$ is the DM interaction vector for the atomic bond $ij$ (in Joule), $\vec{S}_i$ the
atomic moment unit vector, and  the summation is performed on the neighbor pairs $\langle i,j\rangle$.
The direction of $\vec{d}_{ij}$ depends on the type of system considered.
We consider here magnetic ultrathin films, where DMI originates from the interaction with the
high spin-orbit heavy metal of the adjacent layer~\cite{fert1980,fert1990,crepieux1998}.
In this case, for isotropic films $\vec{d}_{ij}$ is $d\:\vec{u}_{ij}\times\hat{z}$\cite{heide2008,heinze2011,fert1990,fert2013,sampaio2013,thiaville2012,noteEPL}, where
$\vec{u}_{ij}$ is the unit vector between sites $i$ and $j$ and $\hat{z}$ is the direction normal to the film oriented from the high spin-orbit layer to the magnetic ultrathin film.
In the micromagnetic framework, the hypothesis that the atomic spin direction evolves slowly at the atomic
scale allows building a continuous form for the DMI. As we consider films that are thinner than any micromagnetic length scale, variations along the surface normal are neglected so that, even if DMI originates from the interfaces, we consider a uniform average value along the film thickness.
Given $\vec{m}(\vec{r})$ the magnetization direction at position $\vec{r}$, the DMI energy reads\cite{bogdanov1989,bogdanov1994}
\begin{equation}\begin{split}\label{eq_energyDM}
    E_{\mathrm{DM}} = t \iint D\left[\left(m_x\frac{\partial m_z}{\partial x}-
m_z\frac{\partial m_x}{\partial x}\right)+\right.\\
    \left.\left(m_y\frac{\partial m_z}{\partial y}-
m_z\frac{\partial m_y}{\partial y}\right)\right]d^2\vec{r}
\end{split}\end{equation}
where $D$ is the continuous effective DMI constant, in J/m$^2$.
The link between $D$ and $d$ depends on the type of lattice, but scales as $1/at$ ($a$ being the lattice constant
and $t$ the film thickness).
The $1/t$ scaling is due to the assuption of interface induced DMI.
We obtain $D = d/at=d/Na^2$ for a simple cubic lattice
oriented along the (001) direction and $D=d\sqrt{3}/at=3d/Na^2\sqrt{2}$ for a face centered cubic lattice
oriented along the (111) direction ($N$ is the number of atomic planes in the film).
For example, given the value from the literature for 1 monolayer of Fe on Ir(111)\cite{heinze2011} $d = -1.8$~meV
and $a=2.715$~\AA, we find $D = -8.3$~mJ/m$^2$ for $N=1$. Note also that, although Eq.~\ref{eq_energyDM} has been derived from a simple first neighbor description, it remains valid for a more complex formulation, as long as the system is isotropic. In such a case, only the link between $D$ and $d$ is modified.


DMI needs to be included together with the other micromagnetic energies so that the exchange energy density $A\left[\left(\frac{\partial\vec{m}}{\partial x}\right)^2+(\frac{\partial\vec{m}}{\partial y})^2\right]$ and anisotropy energy density $-K \left(\vec{m}.\hat{z}\right)^2$ are added to Eq.\ref{eq_energyDM} ($A$ being the micromagnetic exchange constant and $K$ the anisotropy constant).
In this paper, we consider the case with a perpendicular easy axis ($K>0$).
In order to provide exact solutions which can be compared to numerical calculations, we do not consider dipolar coupling here so that $K$ can be seen
as an effective anisotropy constant, which takes into account the shape anisotropy ($K = K_{MC} - \frac{1}{2}\mu_0M_S^2$, with $K_{MC}$ the magnetocrystalline anisotropy and $M_S$ the spontaneous magnetization).
 This approximation is justified by the fact that we are interested in ultrathin films, where dipolar coupling becomes local (shape anisotropy) in the zero thickness limit.\cite{winter1961} See however Fig.~\ref{fig_edge_condition} for a case where full dipolar coupling is included.
As a first approximation, we also do not include any specific edge energies (enhanced edge anisotropy, modified exchange or DMI constant, ...) as usual in continuous magnetism.

For numerical applications, we consider in the following the parameters of Pt/Co/AlOx samples\cite{miron2012} [$A=16$~pJ/m, $K = 510$~kJ/m$^3$ ($\Delta = 5.6$~nm, $D_c = 3.6$~mJ/m$^2$ - see paragraph~\ref{section_DMwall})], thought to be  good candidates to show the importance of DMI\cite{thiaville2012,sampaio2013}. The value of $D$ is varied in order to observe its influence on the micromagnetic configurations.

%


\section{1D case}

We first consider the case where the magnetization direction only changes along the $\hat{x}$ direction.  Such a case has already been considered for an infinite film and the results presented in Sec.~\ref{section_DMwall} and \ref{section_spiral} are already known\cite{dzyalo1964,dzyalo1965,bogdanov1989,heide2008}, but we recall them as they underline the micromagnetic meaning of the parameter $D$ and its associated length $\xi$. Moreover, the results of this 1D model are essential in order to understand results obtained on skyrmions.

Given the fact that, in the case of ultrathin films, $\vec{d}_{ij}$ is orthogonal to $\vec{u}_{ij}$,
the DMI favors rotation in the $(\hat{x},\hat{z})$ plane  with a fixed chirality, so that a single angle $\theta$ is needed to
describe the variation of $\vec{m}(x)$.
Referring $\theta$ to the $\hat{z}$ axis, the total micromagnetic energy density reads
\begin{equation}
\label{eq_energy_1D_continuous}
  E[\theta(x)] = \int_{x_A}^{x_B} \left[A\left(\frac{\partial\theta}{\partial x}\right)^2-
  D\frac{\partial\theta}{\partial x}-K\cos^2\theta\right]dx,
\end{equation}
where $x_A$ and $x_B$ are the boundaries of the sample in the $x$ direction.
We note that, contrarily to the exchange term, the DMI term is chiral so that lowest energy states are
expected for $\partial\theta/\partial x$ of the sign of $D$.\cite{noteEPL}
Using standard variation calculus\cite{brown1963,hubert1998}, it can be shown that the function $\theta(x)$ which minimizes the energy is the solution of the following equations

\begin{subequations}
    \label{eq_mumag1D_all} 
    \begin{eqnarray}
    \frac{d^2\theta}{dx^2} &=& \frac{\sin\theta\cos\theta}{\Delta^2}\mathrm{~~~for~}x_A<x<x_B\label{eq_mumag1D_vol}\\
    \frac{d\theta}{dx}&=&\frac{1}{\xi}\mathrm{~~~for~}x = x_A \mathrm{~or~}x = x_B\label{eq_mumag1D_edge}
    \end{eqnarray}
\end{subequations}
where $\Delta = \sqrt{A/K}$ and $\xi = 2A/D$ are the two characteristic lengths of the problem.
The first one is the well known Bloch wall width parameter~\cite{hubert1998}, while the second one is
\cite{dzyalo1964,dzyalo1965,bogdanov1994,butenko2010}.
By integration of Eq.~(\ref{eq_mumag1D_vol}) we obtain:
\begin{equation}\label{eq_mumag1D_vol_integrated}
    \left(\frac{d\theta}{dx}\right)^2=\frac{C+\sin^2\theta}{\Delta^2}
\end{equation}
where $C$ is an integration constant.

\subsection{Magnetic edge structure and micromagnetic boundary conditions}

Equation (\ref{eq_mumag1D_edge}) needs to be carefully considered.
It corresponds to a condition at the boundary of the sample. Note that no specific micromagnetic energy was considered at the edges so that this is a "natural" boundary condition, that arises from the volume energies. It differs from usual micromagnetism (i.e. without DMI) where it would be $d\theta/dx=0$ in the absence of surface term, or where the edge condition would be due to specific surface energies\cite{thiaville1992,roessler2002,rohart2007,butenko2009}.
A striking consequence is that, in a finite dimension structure with DMI,
the uniform state is never a solution of the micromagnetic problem as soon as $D\neq0$.

For more complex investigations, these boundary conditions have to be implemented in a micromagnetic simulation code, which we have done for two different codes (one homemade, ref.~\onlinecite{code_jacques}, and the public code OOMMF, ref.~\onlinecite{OOMMF}).
Similarly to previous works on micromagnetism \cite{brown1963,miltat1994}, a generalized calculation can be performed for an arbitrary orientation $\vec{n}$ of the edge normal, which leads to the boundary condition
\begin{equation}
    \frac{d\vec{m}}{dn}=\frac{1}{\xi}(\hat{z}\times \vec{n})\times \vec{m}\label{eq:CLgene}.
\end{equation}
This form ensures that the edge magnetization rotates in a plane containing the edge surface normal. Note that the condition does not depend on the definition of normal vector $\vec{n}$ orientation. Similarly, the volume equation, Eq.~(\ref{eq_mumag1D_vol}), can be replaced in a general description by an effective field acting on the local magnetization\cite{brown1963,miltat1994}. The contribution of DMI to this term is
\begin{equation}
    \vec{H}_\mathrm{DM} = \frac{2D}{\mu_0M_S}\left[(\vec{\nabla}.\vec{m})\hat{z}-\vec{\nabla}m_z     \right]
\end{equation}
In order to test the implementation, direct comparisons have been performed between the numerical results and
Eq.~(\ref{eq_mumag1D_all}).

\begin{figure}[ht]
  \includegraphics[width=1.0\columnwidth]{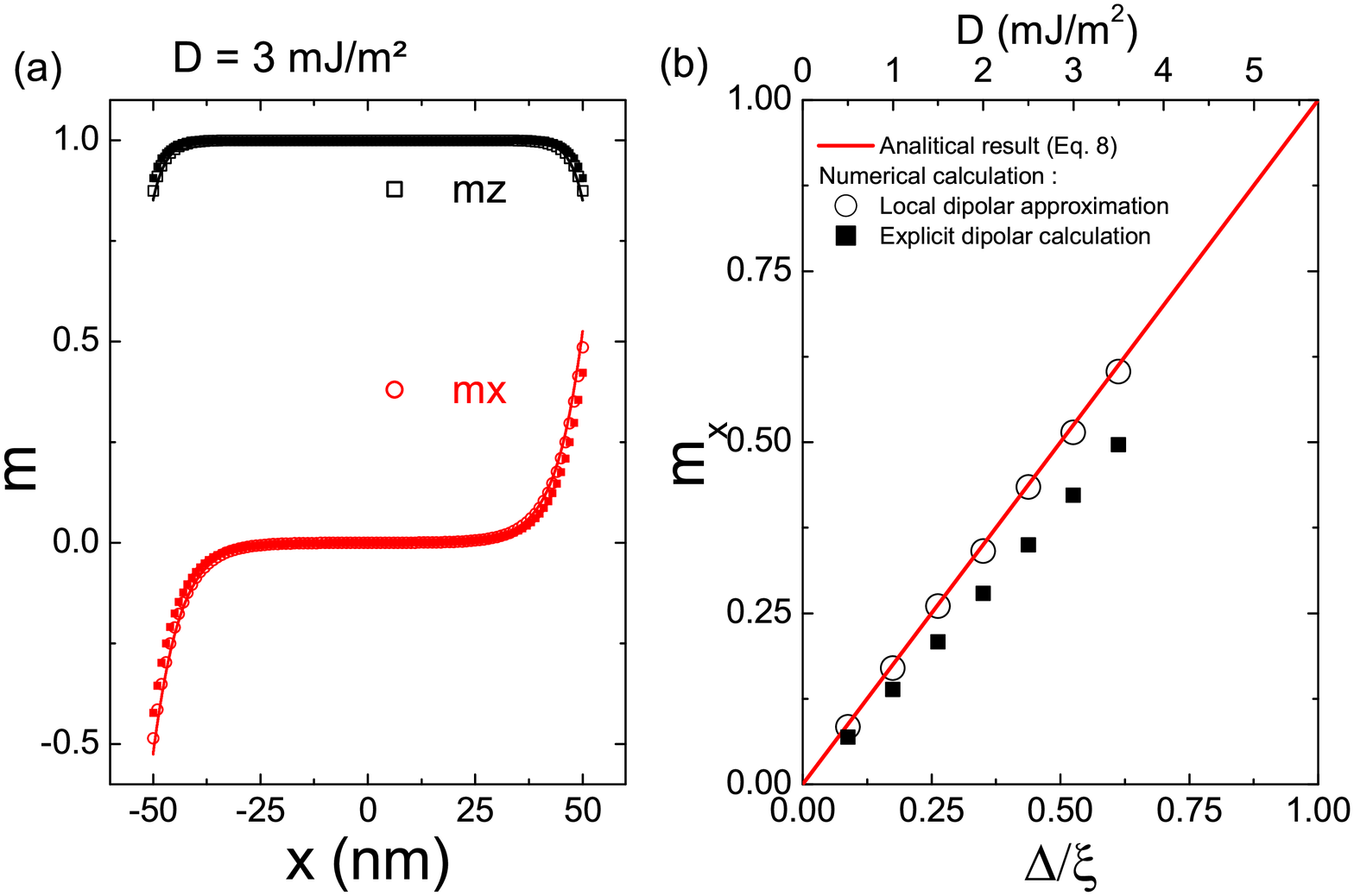}
  \caption{(color online) Magnetization rotation at the edges of an ultrathin film with interface DMI.
(a) Magnetization profile in a stripe infinite in $\hat{y}$ direction and with a 100 nm width in the $\hat{x}$ direction,
with initial magnetization along $\hat{z}$ axis and for $D = 3$~mJ/m$^2$ ($\xi = 10.67$~nm).
(b) Variation of $m_x$ at the structure boundary versus $\Delta/\xi$. The calculation has been stopped at $D = D_c = 3.6$~mJ/m$^2$ as beyond this value, cycloids start to develop in the sample and $C$ in Eq.~\ref{eq_mumag1D_vol_integrated} is not zero. The continuous line is the solution (numerical integration) of Eq.~(\ref{eq_mumag1D_all}) for different strengths of the DMI. In (a) and (b) symbols correspond to numerical calculations: for the open symbols, the local dipolar coupling approximation is used whereas, for the full symbols, the full dipolar energy is included. Note that in (a), both results are hardly distinguishable.}
  \label{fig_edge_condition}
\end{figure}

A particular case arises when the system under consideration has a magneto-crystalline anisotropy sufficiently large
to avoid cycloid configurations in the structure (section~\ref{section_spiral}). Then $C = 0$ in
Eq.~(\ref{eq_mumag1D_vol_integrated}) and, combining Eqs.~(\ref{eq_mumag1D_edge}) and (\ref{eq_mumag1D_vol_integrated}), we find
\begin{equation}
    m_x=\sin\theta=\pm\frac{\Delta}{\xi}
\end{equation}
at the edge of the structure.
The effect of the DMI specific boundary condition is demonstrated in Fig.~\ref{fig_edge_condition},
with a perfect agreement between numerical and analytical calculations.
We observe that, in the center of the structure, the magnetization is uniform and perpendicular to the
film surface.
At the edge, the magnetization tilts in the $(\hat{x},\hat{z})$ plane.
The influence of the edge is felt over a length scale $\Delta$, which is the only characteristic
length in the volume equation, see Eq.~(\ref{eq_mumag1D_vol_integrated}).

In reality, dipolar coupling may slightly modify this result. Indeed, as the magnetization turns out or inward at the edges, magnetic charges are created which limit the magnetization edge tilt. Using  numerical simulation, we have calculated the profile with a full dipolar coupling calculation (for that purpose, we use $M_S = 1.1\times10^6$~A/m and $K_{MC} = 1.27\times10^6$~J/m$^3$ - which corresponds to $K_{eff}= K_{MC}-\frac{1}{2}\mu_0M_S^2= 510$~kJ/m$^3$ as in the previous calculation). The results are plotted in Fig~\ref{fig_edge_condition}. A small reduction of the edge tilt is indeed observed but the overall shape of the magnetization profile is not dramatically modified as anticipated.

\subsection{Dzyaloshinskii domain walls ($D<D_c$)}\label{section_DMwall}

We now consider an infinite system in the $\hat{x}$ direction, and have a closer look at Eq.~(\ref{eq_mumag1D_vol_integrated}).
If $D$ is small enough not to perturb too much the system (domain wall energy remains positive), the integration
constant $C$ must be zero so that no cycloid develops.
It is striking to note that DMI does not appear any longer in this equation.
Eq.~(\ref{eq_mumag1D_vol_integrated}) now has two types of solution.
The first one is uniform (far from the sample edges) with $\theta = 0$ or $\pi$.
The second one corresponds to a domain wall~\cite{hubert1998} with
\begin{equation}\label{eq_theta_wall}
    \theta(x) = 2 \arctan\left[\exp\left(\pm\frac{x-x_0}{\Delta}\right)\right]+n\pi,
\end{equation}
where $x_0$ is the position of the domain wall and $n$ an integer.
The $\pm$ sign determines the chirality of the domain wall and $n$ enables the two types of wall
(from 0 to $\pm\pi$ or from $\pm\pi$ to $\pm2\pi$).
The shape of this domain wall is exactly the same as the Bloch wall obtained without DMI.
Note however that in such calculation with schematic dipolar interaction term, the calculation without
DMI would not impose any condition on the orientation of the rotation~\cite{hubert1998}
(N{\'{e}}el and Bloch walls have the same energy), whereas DMI imposes here a rotation in the
$(\hat{x},\hat{z})$ plane (N{\'{e}}el walls). Note also that, if explicit dipolar interaction were included, small deviations to the Bloch wall profile would occur, due to the magnetic charges created in the wall.\cite{thiaville2012}

The energy of the domain wall can be calculated by injecting Eq.~(\ref{eq_theta_wall}) into
Eq.~(\ref{eq_energy_1D_continuous}).
The integration of the DMI term is straightforward as $\theta$ undergoes a $\pm\pi$ rotation,
giving $\mp\pi D$.
The two other terms are the same as for the wall without DMI\cite{hubert1998}, so that the domain wall energy with DMI is\cite{dzyalo1965,bogdanov1994,heide2008}
\begin{equation}\label{eq_DWenergy}
    \sigma = 4\sqrt{AK}\mp\pi D.
\end{equation}

It is interesting to note that DMI does not change the shape of the 1D domain wall but introduces chirality,
of a sign fixed by that of $D$.
For the most favorable chirality, it lowers the energy.
This property is at the origin of quite interesting dynamic properties of Dzyaloshinskii domain
walls~\cite{thiaville2012}.
The limit of this situation is when $\sigma$ goes to zero.
This defines the critical DMI energy constant $D_c = 4\sqrt{AK}/\pi$.\cite{dzyalo1964,dzyalo1965}
Above it, the domain wall energy is negative so that domain walls proliferate in the sample.
In this case, the integration constant in Eq.~(\ref{eq_mumag1D_vol_integrated}) cannot be zero anymore.

\subsection{Cycloid state ($D>D_c$)}\label{section_spiral}

We now consider a large DMI ($D\geq D_c$).
As domain walls correspond to an energy gain, a cycloid  develops in the sample~\cite{bode2007,ferriani2008},
with $\vec{m}$ rotating in the $(\hat{x},\hat{z})$ plane.
We first consider the simple case where $K=0$ ($D_c = 0$).
In this case, the constant in Eq.~(\ref{eq_mumag1D_vol_integrated}) is determined by minimizing the energy,
integrated over one period $L_0$, to be determined.
This leads to\cite{dzyalo1965}
\begin{subequations}
    \label{eq_cycloideK=0_allequations} 
    \begin{eqnarray}
        \theta(x)&=&\frac{x}{\xi}\label{eq_cycloideK=0}\\
        L_0 &=& 2\pi\xi
    \end{eqnarray}
\end{subequations}
This equation corresponds to a pure cycloid with periodicity $L_0$.
Note that Eq.~\ref{eq_cycloideK=0} is compatible with the edge conditions so that the result is also valid in nanostructures.
This solution gives a physical meaning to the length scale $\xi$ as it describes the
period of cycloids, which develop due to DMI, in a zero anisotropy sample.\cite{dzyalo1964,butenko2010,bogdanov1989,bogdanov1994,bogdanov2002,dzyalo1965}
The larger the intensity of DMI, the shorter the period.

\begin{figure}[ht]
  \includegraphics[width=1.0\columnwidth]{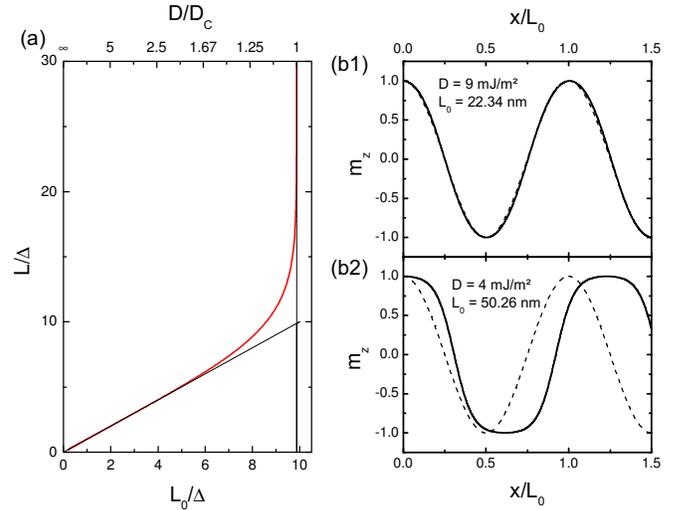}
  \caption{(color online) (a) Variation of the cycloid period $L$ as a function of the anisotropy-free period $L_0$
(result of Eqs.~(\ref{eq_cycloideK}) and (\ref{eq_cycloideE})).
(b) Shape of the cycloid (perpendicular magnetization component $m_z$) in the presence of anisotropy
(the dashed line is the reference cycloid with no anisotropy) for (b1) $D/D_c=2.5$ ($D=9$ mJ/m$^2$, $L_0 = 22.34$~nm) and (b2) $D/D_c=1.1$ ($D=4$ mJ/m$^2$, $L_0 = 50.26$~nm).}
  \label{fig_cycloid}
\end{figure}

If $K\neq0$, a threshold $D_c$ is expected and, as states with $\theta=0$ or $\pi$ are energetically favored,
the pure cycloid should be deformed~\cite{dzyalo1965,degennes1968}.
From Eq.~(\ref{eq_mumag1D_vol_integrated}) we obtain
\begin{equation}
    \frac{d\theta}{\sqrt{C+\sin^2\theta}}=\frac{dx}{\Delta}
\end{equation}
which, integrated over one period $L$, leads to
\begin{equation}\label{eq_cycloideK}
    \frac{L}{4\Delta}=\int_0^{\pi/2}\frac{d\theta}{\sqrt{C+\sin^2\theta}}.
\end{equation}
Integrating the energy over one period and minimizing with respect to $L$ leads to
\begin{equation}\label{eq_cycloideE}
    \frac{D}{D_c}=\frac{\pi^2L}{L_0}=\int_0^{\pi/2}\sqrt{C+\sin^2\theta}\mathrm{~}d\theta
\end{equation}
This last equation determines $C$.
Note that it has a solution only if $D/D_c\geq1$, which validates the previous intuition for the threshold,
based on the domain wall energy.
For $D=D_c$, $C=0$ and the period $L$ diverges.
If $D\gg D_c$, $C$ is large so that $\sin\theta$ can be neglected in Eqs.~(\ref{eq_cycloideK}) and
(\ref{eq_cycloideE}).
This leads to $L\approx L_0$.
In this case, the solution is close to the anisotropy-free solution in Eq.~(\ref{eq_cycloideK=0_allequations}).
Results for any value of $D$ are plotted in Fig.~\ref{fig_cycloid}.

\section{Skyrmions confined in nanodots}

For $D>D_c$, the destabilization of the ferromagnetic state in 2D can lead to the formation of skyrmion
networks~\cite{yu2010,yu2012,heinze2011}.
While calculating such networks is beyond the possibilities of the present formalism, we consider the simple
case of an isolated skyrmion in a circular nanodot of radius $R$, similarly to the model of the vortex studied
by Feldtkeller and Thomas\cite{feldkeller1965}.
The skyrmion being centered in the dot, the circular geometry allows considering radial variations only.
Furthermore, the thin film expression for DMI imposes again a magnetization rotation in the $(\hat{r},\hat{z})$
plane ($\hat{r}$ is the radial unit vector), which produces a hedgehog skyrmion.
The rotation is described by a unique angle $\theta(r)$ referenced from the $\hat{z}$ axis.
The dot energy is
\begin{equation}\label{eq_skyrmion_integral energy}
    \begin{split}
E[\theta(r)] = 2\pi t \int_0^R  \left\{ A \left[\left(\frac{d\theta}{dr}\right)^2+\frac{\sin^2\theta}{r^2}\right] \right.\\
 \left.-D\left[\frac{d\theta}{dr}+\frac{\cos\theta\sin\theta}{r}\right]+K\sin^2\theta\right\} rdr
    \end{split}
\end{equation}
where $t$ is the dot thickness.
A variational calculation leads to the equations for $\theta(r)$:
\begin{subequations}
    \label{eq_sky_allequations} 
    \begin{eqnarray}
        \frac{d^2\theta}{dr^2}&=&-\frac{1}{r}\frac{d\theta}{dr}+\frac{\sin2\theta}{2}\left(\frac{1}{r^2}+\frac{1}{\Delta^2}\right)+\frac{2\sin^2\theta}{\xi r}\label{eq_sky_vol}\\
        \frac{d\theta}{dr}&=&\frac{1}{\xi}\mathrm{~~~for~} r=R\label{eq_sky_edge}
    \end{eqnarray}
\end{subequations}
We note that the edge condition Eq.~(\ref{eq_sky_edge}) is equivalent to that found for the 1D case.
Equation~(\ref{eq_sky_vol}) describes the variation of $\theta$ in the dot. Its solutions have been extensively studied in the case of infinite thin films.\cite{bogdanov1989,bogdanov1994,bogdanov2001,rossler2006,ezawa2010,kiselev2011prl,kiselev2011}
It has no trivial solution respecting the edge condition. In particular, the
uniform state is no more a solution of the problem as soon as $D\ne0$, in analogy with the 1D case.
It has to be integrated numerically with initial value $\theta(r=0)=0$.
The initial value for $d\theta/dr(r=0)$ is adjusted so as to fulfill the boundary condition (shooting method).

For $D = 0$, only one solution is found, the uniform state.
Indeed, when no magnetic field is applied and in the absence of dipolar coupling, no energy can stabilize a
reversed domain (magnetic bubble) in the dot.
When $D$ increases, this uniform solution is slightly modified to fulfill the boundary condition.
We further note that chirality also appears: for this solution, $d\theta/dr$ is of the sign of $D$.

\begin{figure}[ht]
  \includegraphics[width=1.0\columnwidth]{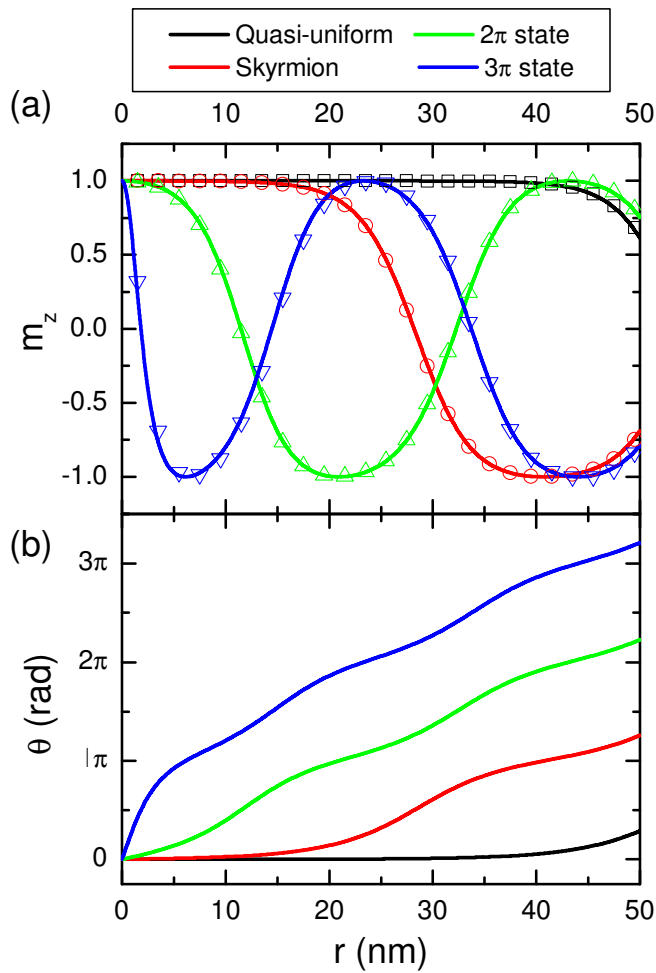}
  \caption{(color online) Results of numerical integration of Eq.~(\ref{eq_sky_allequations})
  for a 100~nm diameter nanodot with $D = 4.5$~mJ/m$^2$ ($D/D_c = 1.2$).
The open symbols are the results of full numerical calculations (with local dipolar energy approximation\cite{winter1961}), given for comparison.
In (b), the variation of $\theta$ shows the chirality imposed by DMI in the micromagnetic configuration.
For this set of parameters, three solutions are found: quasi-uniform (black), skyrmion (red) and $2\pi$ (green)
and $3\pi$ (blue) rotation states.}
  \label{fig_skirmion_config}
\end{figure}

Other solutions also exist.
An example is given in Fig.~\ref{fig_skirmion_config} for $D = 4.5$~mJ/m$^2$.
Four solutions have been found: the uniform one, a skyrmion ($\pi$ rotation) and two other solutions with
larger magnetization rotation ($2\pi$ and $3\pi$ rotation).
In order to test the 2D micromagnetic solvers, simulations have been performed and compared
with these results, as shown in Fig.~\ref{fig_skirmion_config}(a).
Each state is reproduced with a perfect agreement (for the configuration as well as for the energy),
when the energy minimization is started from an initial configuration close enough to the targeted one.

\begin{figure}[ht]
  \includegraphics[width=1.0\columnwidth]{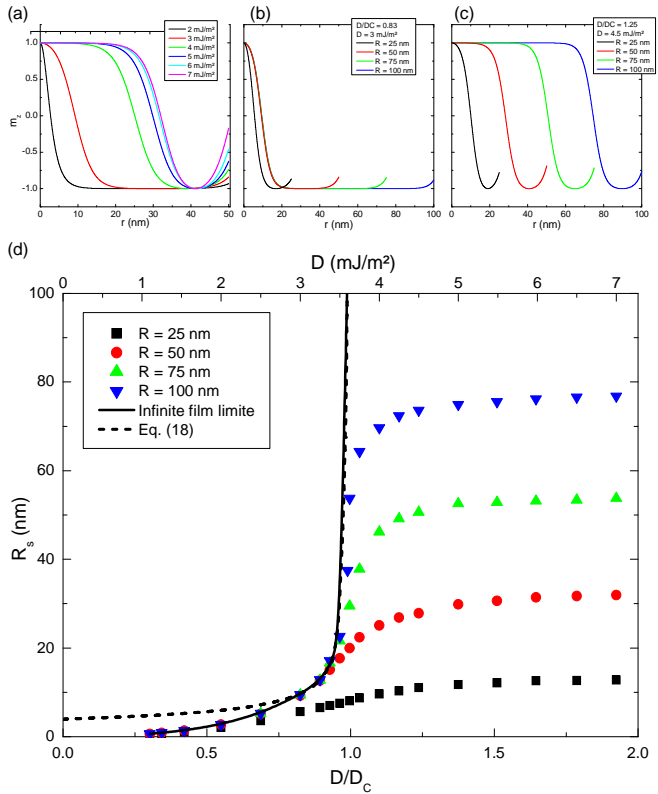}
  \caption{(color online) (a) Variation of the skyrmion profile for different values of $D$, for a 100~nm diameter nanodot.
(d) and (c) Variation of the skyrmion profile versus the dot radius for $D = 3$ and 4.5~mJ/m$^2$
($D/D_c = 0.83$ and 1.25) respectively.
In (c) the skyrmion radius is independent of the dot radius, except for very small radius which
compresses the skyrmion. Note that all these profiles, although $\theta(x)$ is not represented, it corresponds to a monotonic increasing function as in Fig~\ref{fig_skirmion_config}(b), thus to chiral solutions.
(d) Variation of the skyrmion core radius $R_s$ versus $D$ for different dot radius. The radius is defined at the $m_z = 0$ line.
The line is the solution for an infinite thin film and the dotted line is the approximate solution described in the text (Eq.~\ref{eq_sky_diameter}).}
  \label{fig_skirmions}
\end{figure}

The skyrmion solution is similar to a bubble centered in the dot so that the center and the boundary have
opposite magnetization. However, the stabilization of this state is given by DMI only, whereas bubbles are stabilized by external field and/or dipolar coupling\cite{ezawa2010,kiselev2011prl}.
Moreover, we note that this state is different from usual bubbles as the magnetization rotation is chiral,
with a $d\theta/dr$ sign imposed by $D$. The magnetization rotation is not progressive along the radius but
occurs in a narrow range of radius like for a domain wall, and the minimization of anisotropy energy imposes that this transition
occurs on a length scale of $\Delta$. The skyrmion core radius $R_s$ (the line with $m_z = 0$) is mainly controlled by the DMI and increases with $D$ (see Fig.~\ref{fig_skirmions}(a)): as $D$ lowers the domain wall energy, the skyrmion expands to larger diameters when $D$ is large.

To discuss the results, we first consider a single skyrmion, represented as a bubble of radius $R_s$, in an infinite film. Two ranges have to be considered, according to the value of $D$ compared to $D_c$. For $D<D_c$, the domain wall energy $\sigma(D)$, as described in Eq.~(\ref{eq_DWenergy}), is positive so that the skyrmion radius should be zero. However, the domain wall is circular so that a curvature energy cost needs to be included. This term arises from the terms $A\sin^2\theta dr/r$ and $D\cos\theta\sin\theta dr$ in Eq.~(\ref{eq_skyrmion_integral energy}),
which do not appear in the 1D case (Eq.~(\ref{eq_energy_1D_continuous})). As for a domain wall, $\sin\theta\neq0$ only for $r\approx R_S$, if $R_S\gg\Delta$ the variation of $r$ can be neglected in the integral. Using the 1D solution for $\theta(r)$ (Eq.~(\ref{eq_theta_wall})), the skyrmion energy is then
\begin{equation}
    E_s \approx 2\pi R_s t \sigma(D) + \frac{4\pi t A\Delta}{R_s}
\end{equation}
The first term is the domain wall energy cost, the second one the curvature energy cost. The minimisation of this equation gives the skyrmion equilibrium size
\begin{equation}\label{eq_sky_diameter}
    R_s \approx \frac{\Delta}{\sqrt{2(1-D/D_c)}}.
\end{equation}
This solution is plotted as a dotted line in Fig~\ref{fig_skirmions}(d). When $D$ tends toward $D_c$ the skyrmion radius diverges. For small $D$, the radius is small compared to $\Delta$, so that Eq.~(\ref{eq_sky_diameter}) cannot be used; numerical calculations show that $R_s$ goes to zero, as demonstrated previously\cite{kiselev2011}. This type of skyrmions are soliton solutions and have been called isolated skyrmions\cite{kiselev2011}. Note that for the smallest $D$, the skyrmion radius is so small that the magnetization profile is close to an arrow shape\cite{kiselev2011} rather than to that of a magnetic bubble. However, the transition from one shape to the other is continuous in $D$ so that no strict semantic difference can be made between the two shapes, which both are skyrmions. In the second range, for $D>D_c$, the domain wall energy being negative, the previous description does not hold; in infinite films skyrmions\cite{yu2010,yu2012,heinze2011} or cycloids\cite{bode2007} should proliferate, as described previously.

\begin{figure}[ht]
  \includegraphics[width=1.0\columnwidth]{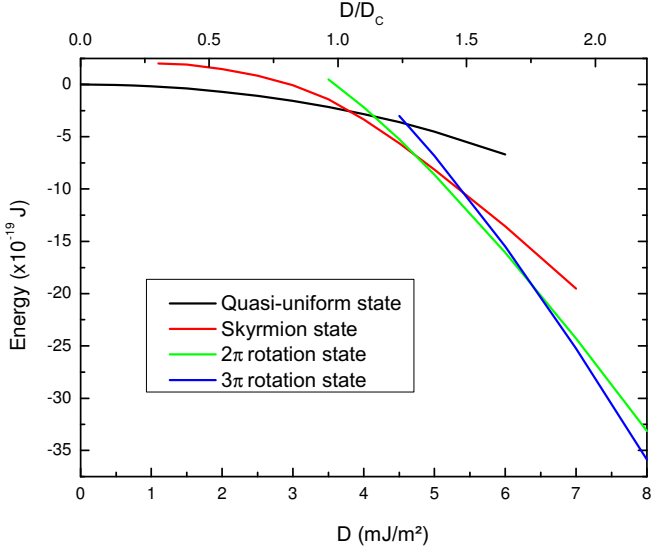}
  \caption{(color online) Variation of the energy of the different states versus $D$, in a 100 nm diameter dot. Note that each line does not cover the full explored $D$ range, as we only plot the solution where a (meta)stable solution has been found. 
  }
  \label{fig_skyE}
\end{figure}

In nanostructures, the situation is rather different as edges play a major role. For the smallest $D$, we found that the skyrmion diameter is independent on the dot diameter and coincides with the infinite film solution (see Fig~\ref{fig_skirmions}(b)). These skyrmions are so small that their shape is not impacted by the edge. For $D\sim D_c$, we do not observe the divergence of the skyrmion diameter and the transition across $D_c$ looks rather continuous. These skyrmions are in fact confined in the dot which limits the diameter increase. Moreover, for $D>D_c$ and if the dot diameter is not too large compared to the cycloid period $L$ (see section~\ref{section_spiral}), a single skyrmion can be isolated in the dot. This sheds light on the important role of the edges, which limit the expansion of the skyrmions. We have identified two main aspects of this confinement. First, for $D>D_c$, the negative domain wall energy means that nothing is expected to limit the growth of an isolated skyrmion. However, in a nanostructure, unlimited increase of the skyrmion radius would let the domain wall move out of the structure, which would turn the dot in the uniform state, with a higher energy (Fig.~\ref{fig_skyE}). This contradiction  proves that necessarily, the edge must limit the growth of the skyrmion and provides a confinement. The skyrmion radius is then fixed by the dot radius (see Fig~\ref{fig_skirmions}(c)). Beyond this, another mechanism also needs to be taken into account, as for $D<D_c$, the domain wall energy being positive, the previous reasoning does not hold. Indeed, if the skyrmion radius increases, as soon as the predicted radius (Eq.~\ref{eq_sky_diameter}) is larger than the dot radius, the dot would turn into the uniform state, with a lower energy (Fig.~\ref{fig_skyE}). The fact that these metastable skyrmions exist even for $D$ close to $D_c$ is the signature of an other confinement energy. It is due to the edge tilting previously described: having the same chirality as the skyrmion, it provides a topological barrier and limits the skyrmion diameter increase. Note the importance of this barrier has been observed in a previous study\cite{sampaio2013} where metastable skyrmions were moved in a track using spin-transfer torque and where it was observed that the edge repels the skyrmions.

In this study, we have considered only a local dipolar coupling  due to the ultrathin film character needed to observe interface induced DMI effects. However, in another study using purely numerical calculations\cite{sampaio2013}, similar results have been obtained with a true dipolar energy calculation, which proves that most of the physics can be captured without the need for sophisticated arguments on this rather complicated energy term.

Other solutions, with more magnetization rotation along the radius also appear. Note that such solutions have been recently observed in skyrmion networks in infinite films.\cite{yu2012}
This is similar to the problem of the cycloid, so that the  length scale is again $L_0$. 
Depending on the value of $D$ they can be more or less stable than the skyrmion.
In the example of Fig.~\ref{fig_skirmion_config}, the third solution with $2\pi$ rotation has an energy
slightly higher than the skyrmion state.
Indeed the dot radius being $R \approx 2L_0$, it seems reasonable to obtain more magnetization rotation.
Finally, the last solution with $3\pi$ rotation is quite unfavorable.
When $D$ is changed the energy of each state changes.
In Fig.~\ref{fig_skyE}, we plot the energy of each state versus $D$.
The four states described previously are not necessarily found for each $D$.
It is interesting to note that the quasi uniform state no longer exists as a metastable state above $D \approx 6$~mJ/m$^2$, and that
the skyrmion exists as metastable state between $\approx1.1$ and $\approx7$~mJ/m$^2$.
However, in the absence of thermal excitation, it becomes more stable than the quasi uniform state as soon as $D\gtrsim D_c$.
As expected, considering the absolute minimum, larger $D$ favors larger  spin rotation so that $n\pi$ solutions (with $n>3$) are expected for  $D$ larger than the explored range.

%
%

\section{Conclusion}

In conclusion, we have considered the effect of the Dzyaloshinskii-Moriya interaction on the micromagnetic configuration in nanostructures, made of ultrathin magnetic films.
One of the most striking effects is the modification of boundary conditions at the edge of nanostructures,
which tilts the edge moments.

The formalism has been applied to describe confined skyrmions in nanodots. The results show that edges are essential to understand such a situation as they provide a confinement and limit the skyrmion expansion. This confinement is rather important for future development of skyrmions-based memories\cite{fert2013} and should deserve further studies in order to be quantitatively understood.



\section*{Appendix}

\subsection{Generalization to other forms for DMI}

We have limited ourselves to the DMI form for ultrathin films.
Much experimental work has also been performed on bulk materials lacking inversion symmetry, belonging to the $D_n$ symmetry group.\cite{yu2010,yu2012} In these, DMI is homogeneous in the volume and $\vec{d}_{ij}=d\vec{u}_{ij}$.~\cite{moriya1960}
For a thin film where magnetization direction variation along the film normal can be neglected,
the continuous DMI energy becomes\cite{bogdanov1989,bogdanov1994}
\begin{equation}\begin{split}
    E_{\mathrm{DM}} = \iiint D\left[\left(m_y\frac{\partial m_z}{\partial x}-m_z\frac{\partial m_y}{\partial x}\right)-\right.\\
    \left.\left(m_x\frac{\partial m_z}{\partial y}-m_z\frac{\partial m_x}{\partial y}\right)\right]d^3\vec{r}
\end{split}\end{equation}
For a 1D system, this interaction favors spin rotation in the $(\hat{y},\hat{z})$ plane
(which means Bloch walls, spirals and vortex-type skyrmions for the different cases considered above),
so that $\theta$ has to be defined in this plane.
In this case, all other equations remain the same, in particular the boundary condition in
Eq.~(\ref{eq_mumag1D_edge}).
Only the general form of the boundary condition for this form of DMI is modified, though the derivation follows the same procedure:
\begin{equation}
d\vec{m}/dn=(\vec{m}\times\vec{n})/\xi.
\end{equation}
Compared to the boundary condition in Eq.~\ref{eq:CLgene}, this one ensures that the edge magnetization rotates in a plane parallel to the edge surface.

\subsection{Extension to thicker samples}

In this paper, we considered the case of DMI ultrathin films with interface DMI. As the sample considered is thinner than $\Delta$ and $\xi$, we assumed a uniform effective DMI constant across the  thickness. This assumption does not hold for thicker samples. In these, the DMI is expressed as a surface term, with $D_{S,i}$ the interface DMI constant (in J/m) where $i$ accounts for the bottom and top interfaces. The micromagnetic energy, limited to DMI and exchange, reads:
\begin{equation}\begin{split}
    E = \sum_iD_{S,i}\iint \left[\left(m_x\frac{\partial m_z}{\partial x}-m_z\frac{\partial m_x}{\partial x}\right)+\right.\\
    \left.\left(m_y\frac{\partial m_z}{\partial y}-m_z\frac{\partial m_y}{\partial y}\right)\right]d^2\vec{r}
    +A\iiint \left(\vec{\nabla}\vec{m}\right)^2d^3\vec{r}
\end{split}\end{equation}
where the surface integral is performed at the interfaces only (assumed normal to $\hat{z}$). Using variational calculation\cite{brown1963,miltat1994} we extract  interface conditions
\begin{subequations}
    \label{} 
    \begin{eqnarray}
        \frac{\partial m_x}{\partial z} = \varepsilon_i\frac{D_{S,i}}{A}\frac{\partial m_z}{\partial x} \\
        \frac{\partial m_y}{\partial z} = \varepsilon_i\frac{D_{S,i}}{A}\frac{\partial m_z}{\partial y}\\
        \frac{\partial m_z}{\partial z} = -\varepsilon_i\frac{D_{S,i}}{A}\left(\frac{\partial m_x}{\partial x}+\frac{\partial m_y}{\partial y}\right)
    \end{eqnarray}
\end{subequations}
with $\varepsilon_\mathrm{bottom} = 1$ and  $\varepsilon_\mathrm{top} = -1$ respectively for the bottom and top interfaces. Note that the boundary conditions have opposite signs on both interfaces. However, for a symmetric stacking (same high spin-orbit non magnetic layer at the bottom and top interfaces), $D_{S,\mathrm{top}} = -D_{S,\mathrm{bottom}}$. Indeed, in the atomic formulation, DMI is proportional to $(\vec{u}\times\hat{z})$, $\hat{z}$ being oriented from the high-spin orbit layer to the magnetic layer, thus opposite for both interfaces.\cite{fert1990} As a consequence, magnetization is bend the same way (i.e. with the same chirality) at both interfaces.

While in such situation, DMI should not be sufficient to destabilize the ferromagnetic state, such boundary condition should modify the structure of domain walls. Indeed, in the volume, Bloch rotation is expected and, at the ferromagnetic film surfaces, N\'eel rotation is expected, with opposite chirality for bottom and top interfaces. This effect, which is purely related to DMI, should add to similar effects due to dipolar coupling.\cite{hubert1998}

\acknowledgments{We thank J. Miltat for the critical reading of the manuscript and acknowledge stimulating discussions with J. Miltat, L. Buda-Prejbeanu, J. Sampaio, V. Cros and A. Fert.
This work was supported by the Agence Nationale de la Recherche, project ANR 11 BS10 008 ESPERADO.}

\bibliographystyle{apsrev}
\bibliography{biblio}

\begin{thebibliography}{43}
\bibitem[{not()}]{noteEPL}
\bibinfo{note}{Note that compared to our first study in
  ref.~\onlinecite{thiaville2012}, we use the opposite convention for the
  product, so that positive $D$ implies that $d\theta/dx>0$ is favored by DMI.}
\end{thebibliography}

\end{document}